\documentclass[
 reprint,
 superscriptaddress,
 amsmath,amssymb,
 aps,
 twocolumn,
]{wlscirep}

\usepackage[english]{babel}

\usepackage{pdftexcmds}
\usepackage{url}
\usepackage{graphicx}% Include figure files
\usepackage{dcolumn}% Align table columns on decimal point
\usepackage{bm}% bold math
\usepackage{hyperref}% add hypertext capabilities
\usepackage{comment}
\usepackage{multirow}
\usepackage{bbold}
\usepackage{physics}

\usepackage[autostyle]{csquotes}
\MakeOuterQuote{"}

\usepackage{xcolor}

\newcommand{\revision}[1]{{\color{black} #1 }} % color

\newcommand{\nn}{\nonumber}
\newcommand{\ms}[1]{\mbox{\scriptsize #1}}
\newcommand{\msi}[1]{\mbox{\scriptsize \textit{#1}}}

\begin{document}

\title{Microwave single-photon detection using a hybrid spin-optomechanical quantum interface}

\author[1,*]{Pratyush Anand}
\author[2]{Ethan G. Arnault}
\author[2,3]{Matthew E. Trusheim}
\author[3,4]{Kurt Jacobs}
\author[1,2+]{Dirk R. Englund} 

\affil[1]{Research Laboratory of Electronics, Massachusetts Institute of Technology, 50 Vassar Street, Cambridge, MA 02139, USA}
\affil[2]{MIT Institute of Soldier Nanotechnology, 500 Technology Square, Cambridge, MA 02139, USA}
\affil[3]{DEVCOM Army Research Laboratory, Adelphi, MD 20783, USA}
\affil[4]{Department of Physics \& Engineering Physics, Tulane University, New Orleans, LA 70118, USA}

\affil[*]{anand43@mit.edu}
\affil[+]{englund@mit.edu}

\date{\today}% It is always \today, today,

\begin{abstract}
Semiconductor single-photon detectors cannot be straightforwardly adapted for the microwave regime, primarily because microwave photons carry far less energy and thus require cryogenic temperatures and specialized architectures. Here, we propose a hybrid spin-optomechanical interface to detect single microwave photons where the microwave photons are coupled to a phononic resonator via piezoelectric actuation. This phononic cavity also acts as a photonic cavity with either a single embedded Silicon-Vacancy (SiV) center in diamond or an ensemble of these centers, bridging optical single-photon detection protocols into the microwave domain. We model the detection process as a communication channel whose capacity is quantified by the mutual information \(I(A;B)\) between the true photon occupancy (A) and the detector outcome (B). Depending on experimentally achievable parameters, simulations predict \(I(A;B)\) in the range \(0.57\,\ln(2)\) to \(0.67\,\ln(2)\), corresponding to true-positive (detection) probabilities above 90\% and false-positive (dark count) probabilities below 10\% per detection interval. These results suggest a viable path to low-noise, high-efficiency single-photon detection at microwave frequencies.

\end{abstract}

%**Abstract (Revised Option 3, with Detection Efficiency and Dark-Count Probability)**  
%**Abstract.** Semiconductor single-photon detectors cannot be straightforwardly adapted for the microwave regime, primarily because microwave photons carry far less energy and thus require cryogenic temperatures and specialized architectures. Here, we propose a spin-optomechanical interface that couples an ensemble of silicon-vacancy centers to a piezoelectric resonator, bridging optical single-photon detection protocols into the microwave domain. We model the detection process as a communication channel whose capacity is quantified by the mutual information \(I_{A,B}\) between the true photon occupancy (A) and the detector outcome (B). Depending on experimentally achievable parameters, simulations predict \(I_{A,B}\) in the range \(0.57\,\ln(2)\) to \(0.67\,\ln(2)\), corresponding to true-positive (detection) probabilities above 90\% and false-positive (dark count) probabilities below 10\% per detection interval. These results suggest a viable path to low-noise, high-efficiency single-photon detection at microwave frequencies.

\maketitle

\section{Introduction}

The advancement of various quantum technologies has enabled quantum optics experiments in the single-photon regime\cite{PhysRevLett.39.691, PhysRevLett.56.58, PhysRevLett.89.187901, Uppu_2020}. This requires efficient single-photon detection schemes in both the optical and microwave (MW) domain. Single-photon detection also provides an important toolkit for quantum information processing and quantum communication\cite{Zoller2005-bs, PhysRevLett.67.661, PhysRevLett.78.3221, Duan_2001, Spiller_2006, Giovannetti_2004, Kimble_2008, Nielsen2012-tx}. There are multiple platforms for single-photon detection in the optical regime\cite{Hadfield2009-te, Eisaman2011-on}. However, in the microwave domain, because of its small energy quanta, the background thermal noise is larger than its optical counterparts, which makes detecting single microwave photons challenging. There have been recent developments in nearly quantum-limited amplification\cite{bergeal2009analog, Macklin2015-mb} and homodyne measurement to extract microwave photon statistics\cite{Lang2013-el} but efficient single-photon detection still remains a puzzle.

The current state of the art for microwave photon detection comprises circuit-QED-based detectors\cite{Gu2017-di, PhysRevLett.102.173602, PhysRevA.84.063834, balembois2023practical, PhysRevLett.107.217401, Sathyamoorthy2016-ge, Inomata2016-he}, opto-electromechanical detectors\cite{barzanjeh2014microwave}, and quantum dot-based detectors\cite{Cornia2023-ko, Wong2017-hr} with an efficiency in the range $\sim0.5-0.7$ and a dark count rate $\alpha$ of $\sim10^{4}-10^{5} s^{-1}$. There are also proposals based on current-biased Josephson junctions\cite{PhysRevLett.107.217401} which is a destructive readout scheme. Non-destructive microwave photon detection with fidelity around 0.9 has been performed using cascaded transmon qubits coupled to a transmission line resonator\cite{Sathyamoorthy2016-ge}. Recently\cite{Inomata2016-he}, an experimental platform of a flux qubit dispersively coupled to a coplanar waveguide (CPW) with a detection efficiency of 0.66 was realized. Here, they implemented an artificial $\Lambda$-type system using a flux qubit and a $\lambda$/2 resonator. However, the circuit-QED-based detection strategies still suffer from low qubit coherence times, short-range connectivity, low qubit number, and low readout fidelities. 

In order to mitigate these issues, there have been efforts to use solid-state defect platforms\cite{PhysRevB.97.205444, Pingault_2017, Faraon_2011, Crook_2020, radulaski2018quantumphotonicsincorporatingcolor, 10.1063/1.1366367, simmons2023scalablefaulttolerantquantumtechnologies, Abobeih_2018, DOHERTY20131, Neuman2021-to, raniwala2024spinoptomechanicalquantuminterfaceenabled} with higher coherence times approaching 10s of milliseconds\cite{PhysRevLett.119.223602}. For color centers (CC), direct MW single-photon-spin coupling is poor ($\sim$ few Hz\cite{Verd__2009}); therefore simply placing a CC in the vicinity of a microwave cavity does not lead to a strong coupling at the single-photon level. To enhance this coupling, intermediate systems can be used to transduce the microwave mode to a phononic mode\cite{Wu2020}. Specifically, recent work\cite{Li2015-bk, Woodman2023-xk} has proposed that a nitrogen vacancy (NV) center in a diamond waveguide can be electromechanically coupled to a CPW cavity in the presence of a micromagnet reaching a single-photon-spin coupling of $\sim$10 kHz. Here, the detection is based on an optical cavity readout and the MW single-photon-spin coupling is mediated via mechanical dark polaritons.

\begin{figure*}[hbt!]
    \centering
    \includegraphics[width=\textwidth]{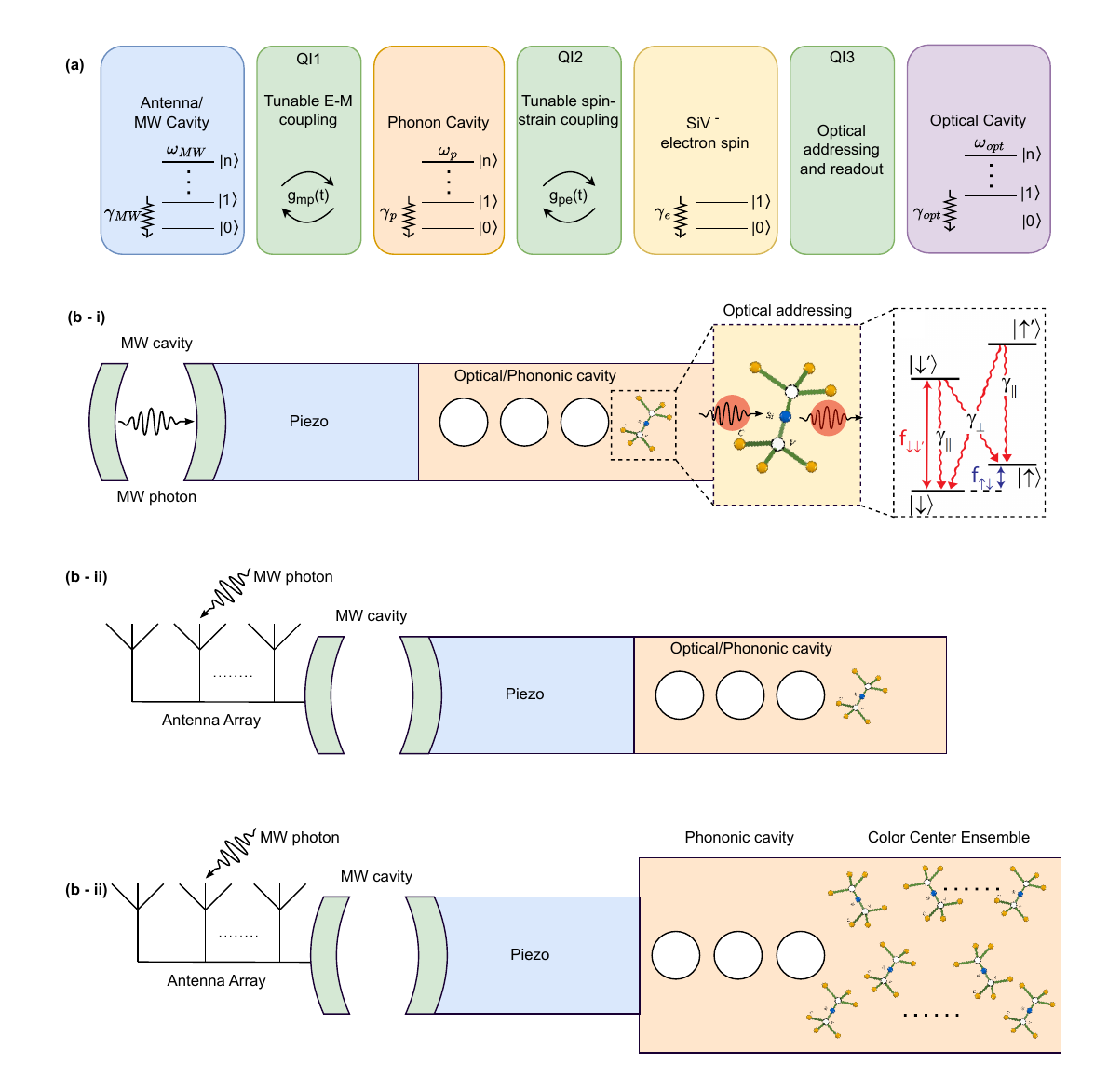}
    \caption{\textbf{(a) A schematic outline of the hybrid spin-optomechanical architecture}. A MW cavity is connected through a piezoelectric quantum interface (QI1) to an optomechanical cavity. The phononic cavity is coupled to the spin-orbit states of the CC(s), which is driven by a piezoelectric interface forming quantum interface 2 (QI2). Further, the optical cavity can be used to form a spin-photon interface (QI3) for optical addressing and single-shot readout of electron spins. \textbf{(b) Architectures for three types of single-photon detector.} (i) Detecting a MW photon present in a cavity, (ii) Detector for a traveling photon for which the shape and arrival time of the wave-packet is known, (iii) Detector for arbitrary incident photons (traditional photon-detector).}
    \label{fig:interface}
\end{figure*}

Inspired by the above use of a mechanical mediator, we propose a detection scheme based on the hybrid spin-optomechanical interface containing a silicon vacancy (SiV$^{-})$ coupled to a microwave resonator via a piezoelectric transducer. Since the energy splitting of group IV CC's depends on local strain\cite{PhysRevB.97.205444}, this allows us to couple the phonon directly with the spin. This platform was recently proposed\cite{Neuman2021-to} as a phononic interface between a superconducting quantum processor and the spin memories of quantum networks, and has been estimated to achieve quantum state transfer with fidelity exceeding 99\% at a microsecond-scale bandwidth. This allows efficient transduction from a microwave photon to microwave phonons in the phononic cavity. The embedded SiV$^{-}$ center allows for AC strain modulation which couples spin and phonon degrees of freedom without the requirement of a micromagnet. This system also allows simultaneous integration into a nanophotonic waveguide due to the electric field sensitivity of the group IV CC\cite{PhysRevB.100.165428}. There are proposals\cite{raniwala2024spinoptomechanicalquantuminterfaceenabled} of even using strain concentrator cavities for SiV$^{-}$, which enhance the single-phonon-spin coupling to $\sim$ 10 MHz. Established methods of spin-readout subsequently allow us to perform single-shot optical readout with fidelity that has been experimentally demonstrated to exceed 99.9\%\cite{Bhaskar2020-ef}. All these conditions provide good motivation to utilize this platform as a single microwave photon detector. 

Here we propose three different architectures for three kinds of single-photon detector. These architectures are shown in Fig.~\ref{fig:interface}b. It is the third architecture, scheme (c), that fulfills the role of the traditional photon counter for traveling-wave photons: it counts incident photons regardless of their arrival times or the shapes of their wavepackets. The first two architectures perform more specialized functions. Scheme (a) detects the presence of a photon in a cavity mode, which would be useful, for example, as part of a quantum processor. It detects a cavity-confined photon using a single CC embedded in an optomechanical cavity. Scheme (b) contains an antenna and thus detects the presence or absence of an incident microwave photon but only so long as the arrival time and shape of the single-photon wave-packet are known in advance. This detector would be useful, for example, in a communication protocol. All three designs work by mapping the quantum state of the MW photon onto the spin state of a CC (or ensemble of CC's), and then apply a measurement to the spin system. We therefore divide the detection protocols into two stages: mapping and readout. 

% This article is structured as follows. We first describe the various mapping and readout schemes used for each of the designs. We then describe in detail the different mapping schemes and provide simulations. After that, we present the two readout schemes. Finally, we combine the two stages and give relevant figures of merit for the three designs. Through these steps we also discuss the Hamiltonians involved, the numerical simulations, and the simulation parameters.

\section{Results}
\subsection{Detection Protocol}
\begin{figure*}[hbt!]
    \centering
    \includegraphics[width=\textwidth]{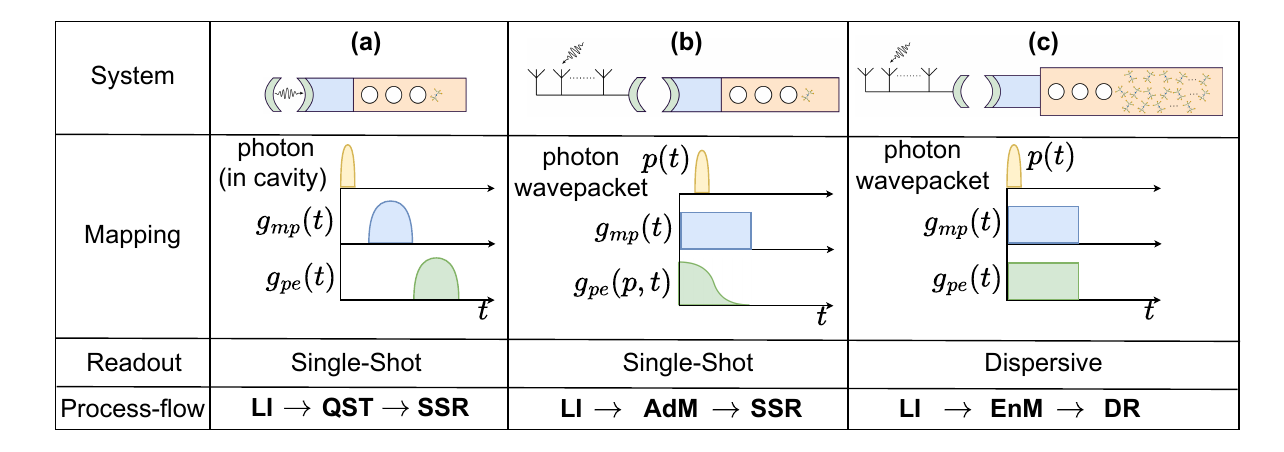}
    \caption{Detection protocols for the three kinds of photon detector. The protocol for each detector has three stages: initialization of the spin(s), transduction/mapping of the input photon into a state of the spin(s), and a readout process for the spin state. Nomenclature: \textbf{LI:} Laser Initialization, \textbf{QST:} Quantum State Transduction, \textbf{SSR:} Single Shot Readout, \textbf{AdM:} Adiabatic Mapping, \textbf{EnM:} Ensemble Mapping, \textbf{DR:} Dispersive Readout.}
    \label{fig:protocols}
\end{figure*}
Fig.~\ref{fig:interface}a shows the overall schematic of the various quantum interfaces involved in the detection protocol. The first step of the scheme is to initialize the spin state using laser. During this process, the laser is parked at the frequency $f_{\uparrow\uparrow^{'}}$ and thus, due to the nonzero cyclicity of optical transitions in SiV, after a sufficient time the spin is initialized to the $\ket{\downarrow}$ state. In addition to that, the number of thermal phonons and thermal photons in the cavities is already very low, as we assume that the system is in thermal equilibrium with a dilution refrigerator at $\sim$ mK temperatures, which corresponds to a thermal photon background, $n_{\ms{th}}\sim0.1$. \revision{We now describe the overall structure of each of the three photon-detector designs shown in Fig.~\ref{fig:interface}}.\\\\
\textbf{Photon detector (a) --- }\revision{Here a MW cavity is coupled to an optomechanical cavity containing a SiV$^{-}$. The purpose is to detect a MW photon present in the cavity. To do so we employ the mapping pulse sequence in Fig.~\ref{fig:protocols}a.} After the initialization, we detune the spin from the phonon modes thereby effectively switching off $g_{\ms{pe}}(t)$, and switching on the tunable E-M coupling $g_{\ms{mp}}(t)$ (QI1). This coupling is mediated via a piezoelectric transducer and acts as a gated detection window. The pulse $g_{\ms{mp}}$ is precharacterized such that it implements a swap operation between the MW and phonon modes. Thus, if there is a MW photon in the cavity before the detection window, it results in a single phonon state in the phononic cavity. After the swap operation, we switch off the coupling $g_{\ms{mp}}(t)$ and switch on the spin-strain coupling $g_{\ms{pe}}(t)$ (QI2) again to perform the swap operation between the electron spin and the phonon mode. After this step, we switch off $g_{\ms{pe}}(t)$ and perform single-shot readout of the electron spin using the spin-photon interface (QI3). In the presence of a MW photon in the cavity, with some fidelity the electron spin will be in the excited state. The fidelity will depend on the swap fidelities and the readout fidelity. After this step, we can reset the system by cooling the spin-phonon modes as done in the first step, and repeat the process.
\revision{
\\\\ \textbf{Photon detector (b) --- }The physical structure of this design is the same as that of (a) except that the microwave cavity is connected to an antenna that can couple with incident traveling-wave photons. As seen in Fig.~\ref{fig:protocols}b, to map the incoming photon wave-packet to the electron spin, this time $g_{\ms{mp}}$ is always on and acts as a detection window. During this time, we drive the pulse $g_{\ms{pe}}(t)$ in a way that is tailored to the shape and arrival time of the single-photon wave-packet $p(t)$. This maximizes the transfer efficiency of the MW photon to the electron spin. The efficiency of this mapping sequence is dependent on the temporal shape, coherence time, and arrival time of the photon wavepacket. We discuss some of these considerations in the numerical simulation section. After mapping the state, we perform the single-shot readout of the CC as in design (a).}
\revision{
\\\\ \textbf{Photon detector (c) --- }The physical system for this design is obtained  by (i) replacing the single CC of design (b) with an ensemble of CCs, and (ii) replacing the optomechanical cavity with a phononic cavity. In this design the quantum state of the MW photon is mapped onto the collective excitation of the spin-ensemble. The couplings, $g_{\ms{mp}}$ and $g_{\ms{pe}}$, are always on as shown in Fig.~\ref{fig:protocols}c. They both act as a detection window and do not need to be tailored to the arrival times or wave-packets of incident photons. An incoming MW photon, upon interacting with the antenna cavity system, begins to be transferred  onto the bright excitation mode of the spin-ensemble via the reversible (Hamiltonian) piezo-electric coupling to the phonons and from the photons via strain to the spins. Due to the inhomogeneity of the spin-ensemble, the bright mode dephases on the timescale of $T^{*}_{2}$. This  irreversibly transfers the bright mode, and thus the incident photon, into the dark modes of the ensemble where it remains until lost on the timescale of $T_{2}$. The irreversible dephasing allows the chain of reversible couplings to act as a tansmission line down which the photon travels to the dark spin modes. After this mapping stage we optically drive the spin ensemble in the dispersive regime, and perform dispersive readout of the collective state. In the next section we discuss the mapping schemes for all three designs in detail.}

\subsubsection{Mapping the photon to a spin-excitation}
In Fig.~\ref{fig:protocols} (bottom line) we show the process flow for the detection schemes for the three systems. For systems (a), (b), and (c), the process of mapping the quantum state of a MW photon to the spin(s) involves, respectively, quantum state transduction, adiabatic mapping, and ensemble mapping.
\newline\newline
\textbf{Quantum State Transduction}\\
The Hamiltonian describing the transduction involving the MW, phonon, and spin degrees of freedom is given by:
\begin{equation}
\label{eq1}
\begin{split}
    \hat{H}_{m-e} &= \hbar\omega_{\ms{mw}}\hat{a}^{\dagger} \hat{a} + \hbar\omega_{\ms{p}}\hat{b}^{\dagger} \hat{b} + \hbar\omega_{\ms{e}}\hat{\sigma}^{\dagger}_{\ms{e}} \hat{\sigma}_{\ms{e}}\\
    &+ \hbar g_{\ms{mp}}(t)(\hat{a}^{\dagger} \hat{b} + \hat{a} \hat{b}^{\dagger}) + \hbar g_{\ms{pe}}(t)(\hat{\sigma}^{\dagger}_{\ms{e}} \hat{b} + \hat{\sigma_{\ms{e}}} \hat{b}^{\dagger})
\end{split}
\end{equation}
Here $\hat{a} (\hat{a}^{\dagger})$ is the MW cavity annihilation (creation) operator, $\hat{b} (\hat{b}^{\dagger})$ is the phononic cavity annihilation (creation) operator, and $\hat{\sigma}_{\ms{e}} (\hat{\sigma}_{\ms{e}}^{\dagger})$ is the electron spin lowering (raising) operator. The frequencies $\omega_{\ms{mw}}$, $\omega_{\ms{p}}$, and $\omega_{\ms{e}}$ correspond to the MW, phonon, and electron-spin qubit, respectively. 
To incorporate losses into the system, we use the Lindblad equation of motion for the density matrix $\hat{\rho}$, and the Lindblad superoperators $\gamma_{\ms{c}_{i}}\mathcal{L}_{\ms{c}_{i}}(\hat{\rho})$:
\begin{equation}
\label{eq2}
\frac{d}{dt}\hat{\rho} = \frac{1}{i\hbar}[\hat{H}_{m-e}, \hat{\rho}] + \sum_{i}\gamma_{\ms{c}_{i}}\mathcal{L}_{\ms{c}_{i}}(\hat{\rho}),
\end{equation}
where
\begin{equation}
\label{eq3}
    \gamma_{\ms{c}_{i}}\mathcal{L}_{\ms{c}_{i}}(\hat{\rho}) =  \frac{\gamma_{\ms{c}_{i}}}{2}\bigg(2\hat{c}_{i}\hat{\rho} \hat{c}_{i}^{\dag} - \{\hat{c}_{i}^{\dag}\hat{c}_{i}, \hat{\rho}\}\bigg),
\end{equation}
with $\hat{c}_{i}$ $\in$ $\{\hat{a}, \hat{b}, \hat{\sigma}_{\ms{e}}\hat{\sigma}_{\ms{e}}^{\dagger}\}$, and $\gamma_{\ms{c}_{i}}$ $\in$ $\{\gamma_{\ms{mw}}, \gamma_{\ms{p}}, \gamma_{\ms{e}}\}$ which represents the decay (or decoherence) rates of the respective modes. The Lindblad superoperators $\mathcal{L}_{\ms{a}}(\hat{\rho})$ and $\mathcal{L}_{\ms{b}}(\hat{\rho})$ describe the $T_{1}$ processes of MW cavity decay and phonon decay, respectively. The super-operator $\mathcal{L}_{\sigma_{\ms{e}}\sigma_{\ms{e}}^{\dagger}}(\hat{\rho})$ describes the $T_{2}$ process of pure dephasing of the electron spin qubit. 

\begin{figure*}[hbt!]
    \centering
    \includegraphics[width=\textwidth]{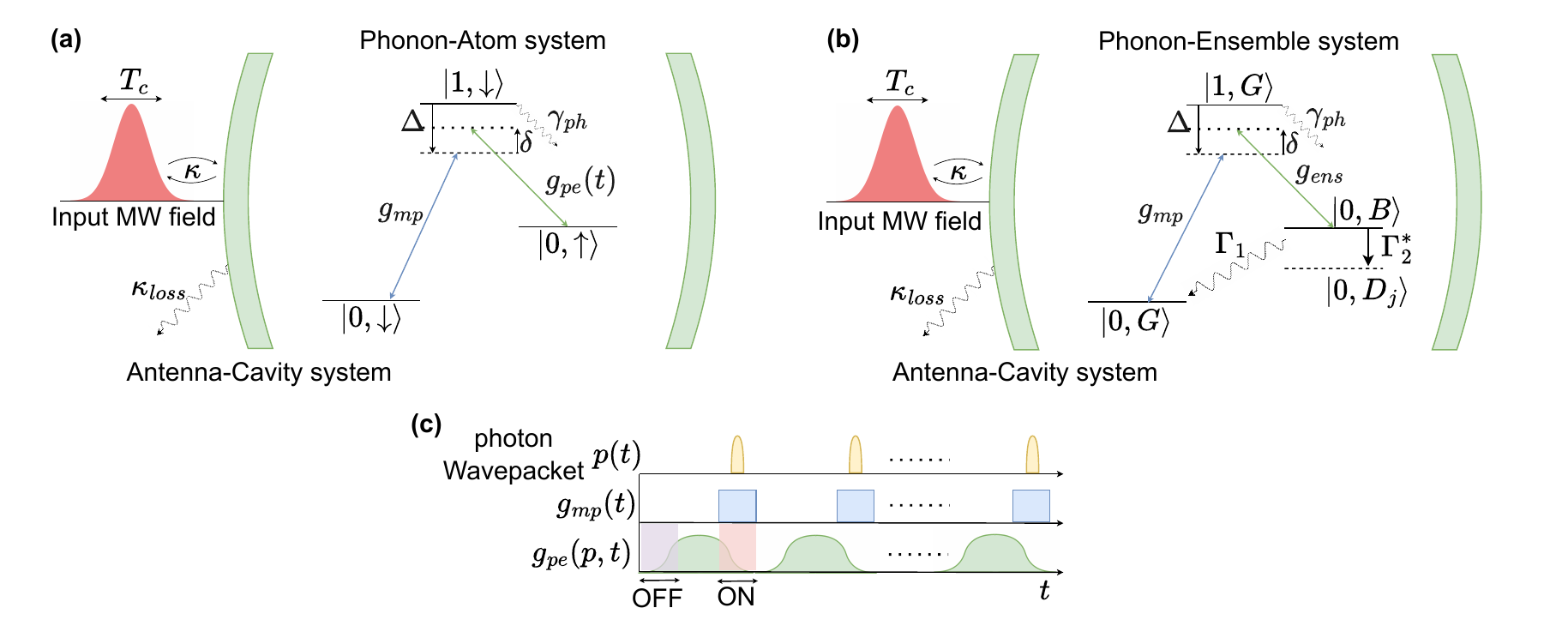}
    \caption{\textbf{(a) Adiabatic Mapping}: in this process an incoming photon causes a change to the state of the CC spin. The transition $\ket{0,\downarrow}\leftrightarrow\ket{1,\downarrow}$ is coupled to the Antenna-MW cavity at rate $g_{\ms{mp}}$, thereby swapping the MW and phonon state. Simultaneously, the time-dependent drive $g_{\ms{pe}}(t)$ couples the 1-fock manifold $\ket{1,\downarrow}\leftrightarrow\ket{0,\uparrow}$.
    \textbf{(b) Ensemble Mapping}: in this process an incoming photon causes a change to the state of the ensemble of CC spins. The transition $\ket{0,G}\leftrightarrow\ket{1,G}$ is coupled to the Antenna-MW cavity at rate  $g_{\ms{mp}}$, thereby swapping MW and phonon state. The effective phonon-spin ensemble coupling $g_{\ms{ens}}$ couples the 1-Fock manifold $\ket{1,G}\leftrightarrow\ket{0,B}$. Due to the spin inhomogeneity, the bright state $\ket{0,B}$ dephases to one of the ensemble dark states $\ket{0,D}$ on the timescale of $T_{2}^{*}$. This results in the irreversible absorption of the photon into a collective dark state of the ensemble. \textbf{(c) Repetition of the adiabatic mapping}: The red shaded area (ON) serves as the detection window. In the purple shaded region (OFF), the drive $g_{\ms{pe}}$ is time-reversed from that in the ON region, and the coupling $g_{\ms{mp}}$ is turned off to prevent emission of then photon back into the MW cavity.}
    \label{fig:Adm}
\end{figure*}

As can be seen from Fig.~\ref{fig:protocols}a, the state-transfer protocol is based on two swap operations mediated by pulses $g_{\ms{mp}}(t)$ and $g_{\ms{pe}}(t)$. From a previous work\cite{Neuman2021-to} on quantum state transduction, in order to avoid high-frequency components, we assume that the couplings have a smooth dependence on time given by
\begin{equation}
\label{eq4}
    g_{\ms{mp}}(t) = g_{\ms{mp}} \sech(2g_{\ms{mp}}(t - \tau_{\ms{mp}}))
\end{equation}
\begin{equation}
\label{eq5}
    g_{\ms{pe}}(t) = g_{\ms{pe}} \sech(2g_{\ms{pe}}(t - \tau_{\ms{pe}})),
\end{equation}
where $g_{\ms{mp}}$, $g_{\ms{pe}}$ are time-independent amplitudes and $\tau_{\ms{mp}}$, $\tau_{\ms{pe}}$ are time delays for the respective pulses. The time delay between pulses, $\Delta\tau_{mpe}$ = $\tau_{\ms{pe}}-\tau_{\ms{mp}}$, can be further adjusted for a given value of the parameters [$\gamma_{\ms{mw}}, \gamma_{\ms{p}}, \gamma_{\ms{e}}, g_{\ms{mp}}, g_{\ms{pe}}$] in order to optimize the state-transfer fidelity $\mathcal{F}$ defined as:
\begin{equation}
   \label{eq6}
    \mathcal{F} = \left| \mbox{Tr} \left[   
      \sqrt{ \sqrt{ \hat{\rho}_{\ms{i}} }  
                    \hat{\rho}_{\ms{f}}
             \sqrt{ \hat{\rho}_{\ms{i}} }  } \right] \right|,
\end{equation}
where $\hat{\rho}_{\ms{i}}$ and $\hat{\rho}_{\ms{f}}$ are, respectively, the density matrices corresponding to the initial state of the MW cavity and the final state of electron spin.
\newline\newline
\textbf{Adiabatic Mapping}\\
For the process of adiabatically mapping the traveling MW photon to the excited state of the atomic spin, we refer to Fig.~\ref{fig:Adm}a, which depicts the joint basis of the phonon mode and the spin qubit. We denote the state in which there are $n$ photons and the spin is the excited (ground) state by $\ket{n,\uparrow}$ ($\ket{n,\downarrow}$).
The $\Lambda$-type transition allows one to perform an effective Raman transition ($\ket{0,\downarrow}\leftrightarrow\ket{0,\uparrow}$) in the presence of two drives, $g_{\ms{mp}}$ and $g_{\ms{pe}}$. We adapt the analytical analysis of the optimal driving from references\cite{Giannelli_2018, Giannelli_2019}. The Hamiltonian describing the coherent process is given by $\hat{H}(t)$ = $\hat{H}_{\ms{fields}}$ + $\hat{H}_{\ms{I}}(t)$, where
\begin{equation}
\label{eq7}
\hat{H}_{\ms{fields}} = \sqrt{\kappa}~\mathcal{E}_{\ms{in}}(t)(\hat{a} + \hat{a}^{\dagger})
\end{equation}
and $\hat{H}_{\ms{I}}(t)$ is given below in Eq.(\ref{eq10}). 
In Eq.(\ref{eq7}) the parameter $\kappa$ is the cavity decay due to coupling of the antenna with the continuum of electromagnetic modes (the incoming photons), whereas the cavity loss to other modes is represented by $\kappa_{\ms{loss}}$. The operators $\hat{a}$ ($\hat{a}^{\dagger}$) correspond to the annihilation (creation) of the cavity mode. For simplicity, we assume that the incoming wavepacket is a weak coherent pulse with temporal shape $\mathcal{E}_{\ms{in}}(t)$ and mean photon number $n$~\footnote{In fact, the equations of motion for a single-photon wave-packet input to a cavity are identical to that for a coherent pulse, in which the wave-packet of the former maps to the amplitude profile of the latter\cite{Heuck2020}.}. While we assume the input profile to be a hyperbolic secant, the formalism is independent of the exact nature of the profile. Our input profile is thus:
\begin{equation}
    \label{eq8}
    \mathcal{E}_{\ms{in}}(t) = \frac{\sqrt{n}}{\sqrt{T}}\sech\bigg(\frac{2t}{T}\bigg),
\end{equation}
where the coherence time $T_{\ms{c}}$ of the wavepacket given by $T_{\ms{c}}=\pi T/4\sqrt{3}$. The input profile satisfies the following relation:
\begin{equation}
    \label{eq9}
    \int^{+\infty}_{-\infty}~|\mathcal{E}_{\ms{in}}(t)|^{2}~\mathrm{dt} = n
\end{equation}
The interaction Hamiltonian in the rotating wave approximation is given by,
\begin{equation}
    \label{eq10}
\begin{split}
   \hat{H}_{\ms{I}}(t) &= \delta\ket{0,\uparrow}\bra{0,\uparrow} - \Delta\ket{1,\downarrow}\bra{1,\downarrow}\\
   &+ g_{\ms{mp}}(\hat{a}^{\dagger}\ket{0,\downarrow}\bra{1,\downarrow} + \hat{a}\ket{1,\downarrow}\bra{0,\downarrow})\\
   &+ g_{\ms{pe}}(t)\ket{1,\downarrow}\bra{0,\uparrow} + g_{\ms{pe}}^{*}(t)\ket{0,\uparrow}\bra{1,\downarrow}, 
\end{split}
\end{equation}
where $\Delta=\omega_{\ms{c}}-\omega_{\ms{ph}}$ is the detuning between the antenna-cavity frequency $\omega_{\ms{c}}$ and phonon-cavity frequency $\omega_{\ms{ph}}$, and $\delta=\omega_{\ms{e}}+\omega_{\ms{o}}-\omega_{\ms{c}}$ is the detuning between spin frequency $\omega_{\ms{e}}$ and antenna-cavity frequency, evaluated using the central frequency $\omega_{\ms{o}}$ of the drive $g_{\ms{pe}}(t)$. The incoherent evolution is simulated using Eq.(\ref{eq2}), where the modified collapse parameters are $c_{i}\in\{\hat{a},\hat{b}, (\ket{0,\uparrow}\bra{0,\uparrow}-\ket{0,\downarrow}\bra{0,\downarrow})\}$, and $\gamma_{\ms{c}_{i}}\in\{\kappa + \kappa_{\ms{loss}},\gamma_{\ms{ph}},\gamma_{\ms{e}}\}$.
The initial state of the system is
% \begin{equation}
%     \label{eq11}
%     \ket{\psi_{0}} = \ket{\psi_{coh}}\otimes\ket{0}_{\ms{c}}\otimes\ket{0}_{\ms{ph}}\otimes\ket{\downarrow}
% \end{equation}
\begin{equation}
    \label{eq11}
    \ket{\psi_{0}} = \ket{0}_{\ms{c}}\otimes\ket{0}_{\ms{ph}}\otimes\ket{\downarrow}
\end{equation}
and the projection operator for the desired target state is
\begin{equation}
    \label{eq12}
    \hat{P}_{T} = \mathbb{1}_{\ms{c}}\otimes\mathbb{1}_{\ms{ph}}\otimes\ket{\uparrow}\bra{\uparrow}
\end{equation}
Here $\mathbb{1}_{\ms{c}}$ and $\mathbb{1}_{\ms{ph}}$ denote, respectively, the identity operators for the MW and phonon modes. The transfer efficiency to map the state is then defined as
\begin{equation}
    \label{eq13}
    \eta(t) = \mathrm{Tr}(\hat{P}_{T}~\hat{\rho}(t)).
\end{equation}
The fidelity of transfer $\nu$ is defined as the ratio of the transfer efficiency and the number of photons impinging on the antenna-cavity system: 
\begin{equation}
    \label{eq14}
    \nu(t) = \frac{\eta(t)}{\int^{t}_{t_1} |\mathcal{E}_{\ms{in}}(\tau)|^{2}\mathrm{~d}\tau}.
\end{equation} 
The time interval for mapping is $[t_1, t_2]$, with  $t_1<t<t_2$, and we assume that the incoming wavepacket interacts with the antenna only in this time interval. Thus, at the completion of the mapping process the transfer fidelity is 
\begin{equation}
    \label{eq15}
    \nu \equiv \nu(t_2) = \eta(t_2)/n . 
\end{equation}
Given our definition of the transfer efficiency, in the limit of a weak coherent pulse, $n \ll 1$, the transfer fidelity $\nu$ approaches the single-photon storage efficiency $\eta_{\ms{sp}}$ as described here\cite{Giannelli_2019}. In\cite{Giannelli_2018, Giannelli_2019}, the authors show that in the adiabatic regime (i.e. $\gamma T_{\ms{c}} C \gg 1$), the optimal drive that maximizes $\nu(t_{2})$ is given by:
\begin{equation}
    \label{eq16}
\begin{split}
    g_{\ms{pe}}^{\msi{opt}}(t)&=\frac{\gamma\left(1+C\right)+\mathrm{i} \Delta}{\sqrt{2 \gamma\left(1+C\right)}} \frac{\mathcal{E}_{\text {in }}(t)}{\sqrt{\int_{t_1}^t\left|\mathcal{E}_{\text{in }}\left(\tau\right)\right|^2 \mathrm{~d} \tau}}\\
    &\times \exp \left(-\mathrm{i} \frac{\Delta}{2 \gamma\left(1+C\right)} \ln \int_{t_1}^t\left|\mathcal{E}_{\text {in }}\left(\tau\right)\right|^2 \mathrm{~d} \tau\right),
\end{split}   
\end{equation}
where $C$ is the effective cooperativity of the system, 
\begin{equation}
    \label{eq17}
C = \frac{g_{\ms{mp}}^2}{\gamma_{\ms{ph}}(\kappa + \kappa_{\ms{loss}})}.
\end{equation}
Below we will discuss the numerical simulation of the optimum adiabatic mapping. Naturally one can continuously repeat this protocol to detect multiple incoming wavepackets as depicted in Fig.~\ref{fig:Adm}c.  
\newline\newline
\textbf{Absorption of the photon by the ensemble}\\
There have been a number of works devoted to mapping the quantum state of a microwave photon wavepacket to an atomic ensemble in the context of spin-ensemble based quantum memories\cite{PhysRevA.88.062324, PhysRevLett.102.083602, PhysRevA.86.063810, PhysRevLett.110.250503}. However, for detection of single-photons, we do not need to retrieve the stored photon. Thus for design (c) we use a similar scheme as those for photon storage with the constraint of efficient retrieval relaxed. 

In Fig.~\ref{fig:Adm}b we show a schematic of the ensemble mapping process. The primary difference from that of mapping to a single spin is that the ensemble dephasing process continually transforms the bright collective state of the ensemble to which the photon is being mapped into the space of \enquote{dark} collective spin states that do not interact with the phonon mode. In this way the spin ensemble acts as a bath that irreversibly transfers the photon into the dark collective states of the ensemble. This irreversible process allows the ensemble to absorb the photon without having to tailor time-varying drives $g_{\ms{mp}}$ and $g_{\ms{pe}}$. This occurs because the decay rate into the spin ensemble is impedance matched to the decay rate of the microwave cavity to ensure that the photon is absorbed by the cavity (and in turn the ensemble) rather than being reflected from it. 

The Hamiltonian for the coherent part of the process is given by $\hat{H}(t)$ = $\hat{H}_{\ms{fields}}$ + $\hat{H}_{\ms{ens}}(t)$, where the description of the wavepacket is the same as in Eqs.(\ref{eq7})-(\ref{eq9}). The other part of the Hamiltonian is given by
\begin{equation}
    \label{eq18}
\begin{split}
    \hat{H}_{\ms{ens}} &= \hbar\omega_{\ms{mw}}\hat{a}^{\dagger}\hat{a} + \hbar\omega_{\ms{ph}}\hat{b}^{\dagger}\hat{b} + \sum^{N}_{j=1}\hbar(\omega_{j}/2)\hat{\sigma}_{z,j}\\
    &+ \hbar g_{\ms{mp}}(\hat{a}^{\dagger}\hat{b} + \hat{a}\hat{b}^{\dagger}) + \sum^{N}_{j=1}\hbar g_{j}(\hat{b}^{\dagger}\hat{\sigma}_{-,j} + \hat{b}\hat{\sigma}_{+,j}) . 
\end{split}    
\end{equation}
Here $\hat{a}$ and $\hat{b}$ are the MW and phononic cavity annihilation operators, $\hat{\sigma}_{x,j}$, with $x = z,\pm$ are the various spin operators for the $j^{\msi{th}}$ spin qubit, and $\omega_{j}$ and $g_{j}$ are, respectively, the splitting frequency and the phonon coupling rate for the $j^{\msi{th}}$ spin qubit. In the limit in which the number of excitations in the spin ensemble is much smaller than the number of spins, $N$, the ensemble effectively behaves as a resonator. Therefore, the Pauli lowering operator ($\hat{\sigma}_{-,j}$) can be replaced by a Bosonic annihilation operator ($\hat{s}_{j}$). In the interaction picture the Hamiltonian becomes 
\begin{equation}
    \label{eq19}
\begin{split}
    \hat{H}_{\ms{ens}} &= \sum^{N}_{j=1}(\delta_{j} \ket{0}\bra{0}\otimes\hat{s}_{j}^{\dagger}\hat{s}_{j} - \Delta_{j} \ket{1}\bra{1}\otimes\hat{s}_{j}\hat{s}_{j}^{\dagger})\\
    &+ \hbar g_{\ms{mp}}(\hat{a}^{\dagger}\hat{b} + \hat{a}\hat{b}^{\dagger}) + \hbar g_{\ms{ens}}(\ket{1,G}\bra{0,B} + \ket{0,B}\bra{1,G}).
\end{split}
\end{equation}
In the collective excitation picture of the ensemble, $\ket{G}$ and $\ket{B}$ correspond to the collective ground state and first bright state respectively, related by the following relations:
\begin{align}
    \label{eq20}
    \ket{B} & = \hat{s}^{\dagger}_{\ms{ens}}\ket{G} , \\
    \label{eq21}
    \hat{s}_{\ms{ens}} & = \sum^{N}_{j=1}g_{j}\hat{s}_{j}/g_{\ms{ens}} , \\
    \label{e22}
    {g}_{\ms{ens}} & = \sqrt{\sum^{N}_{j=1}g^{2}_{j}} = \sqrt{N} \sqrt{\langle g_j^2 \rangle}, 
\end{align}
where $\langle g_j^2 \rangle$ denotes the squared coupling of each of the spins averaged over the ensemble. In the above equations $\hat{s}_{\ms{ens}}$ and $g_{\ms{ens}}$ are the collective excitation operator and collective coupling of the ensemble, respectively, in which the latter is enhanced by $\sqrt{N}$. In the presence of inhomogeneous broadening, which is characterized by the width $\Gamma$ of the distribution of spin resonance frequencies $\rho(\omega_{j})$, the bright mode $\ket{B}$ is not an eigenstate and thus redistributes/decays to the $N-1$ excitation (dark) states $\ket{D_{j}}$ on the timescale of $T^{*}_{2}=\Gamma^{-1}$ (similar to Free Induction Decay). Fig.~\ref{fig:Adm}b shows an effective state-level diagram of the various transitions. This incoherent evolution is simulated using an equation similar to Eq.(\ref{eq2}), where the collapse parameters are given by $c_{i}\in\{\hat{a},\hat{b},\ket{0,D}\bra{0,B}\}$ and $\gamma_{\ms{c}_{i}}\in\{\kappa + \kappa_{\ms{loss}},\gamma_{\ms{ph}},\Gamma\}$. Since, the incoming photon state is mapped to state $\ket{D}$, we define the following projection operator $\hat{P}_{T}$ as
\begin{equation}
    \label{eq23}
    \hat{P}_{T} = \mathbb{1}_{\ms{c}}\otimes\mathbb{1}_{\ms{ph}}\otimes\ket{D}\bra{D}.
\end{equation}
We use the same definitions for the transfer efficiency ($\eta$) and fidelity ($\nu$) as given in Eqs.(\ref{eq11})-(\ref{eq15}). After mapping the state onto the dark state, we propose to use a dispersive readout scheme which we discuss below.

\subsubsection{Reading out the spin state}

\begin{figure*}[hbt!]
    \centering
    \includegraphics[width=\textwidth]{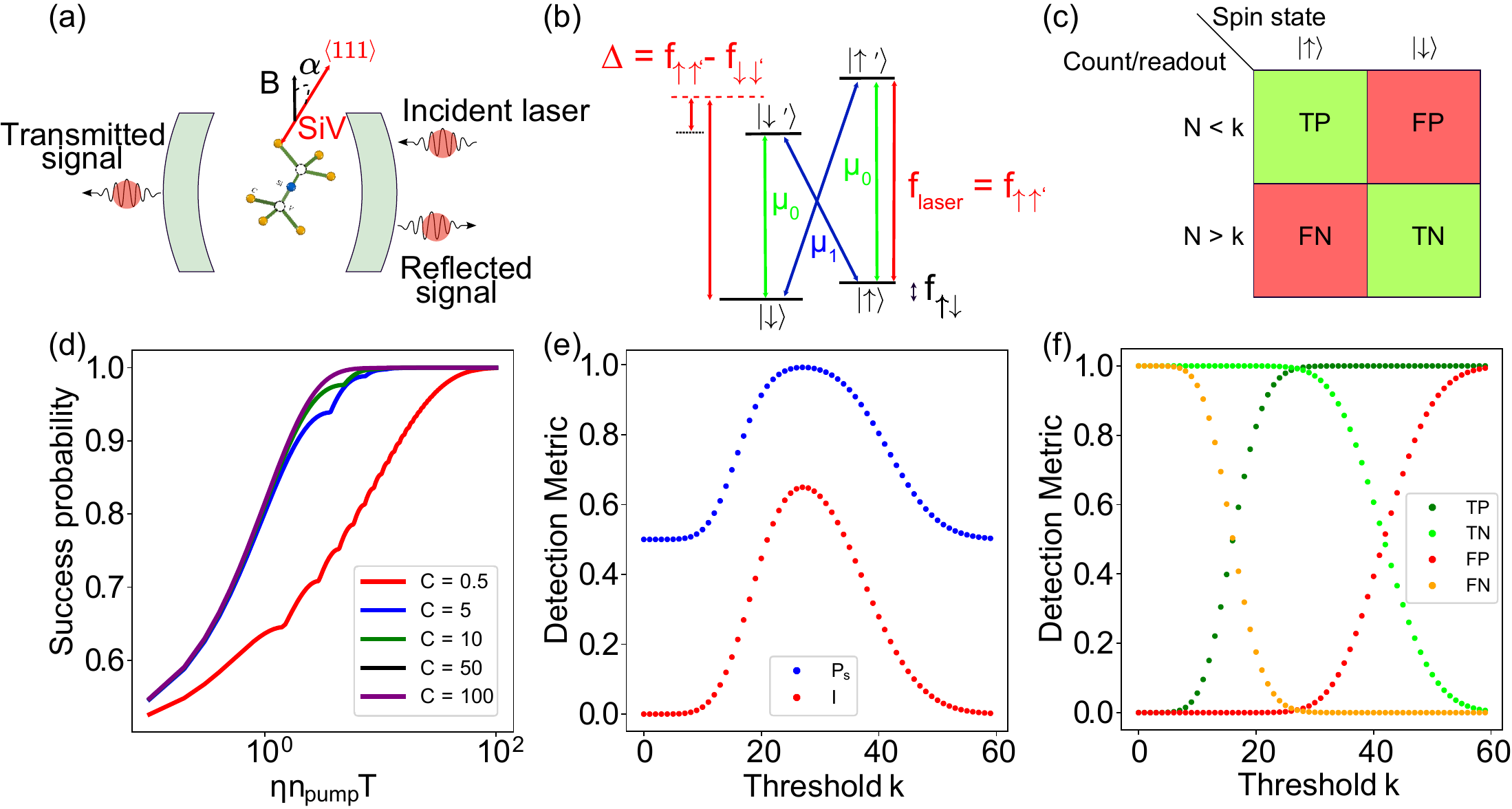}
    \caption{\textbf{(a)} Cavity-enhanced single-shot readout of spins. \textbf{(b)} Energy level structure of the SiV$^-$ with laser probing of the red transitions, green representing the spin-conserving transitions, blue representing the spin-flipping transitions. \textbf{(c)} \textbf{Confusion matrix representation of the readout scheme}. \textbf{\mbox{TP}}: True Positive, \textbf{\mbox{TN}}: True Negative, \textbf{\mbox{FP}}: False Positive, \textbf{\mbox{FN}}: False Negative. \textbf{(d)} Success probability of single-shot readout as a function of probe laser duration $T$ and cooperativity $C$. Variation of the detection metrics \textbf{(e)} $P_{\ms{s}}$, $I(X;Y)$, \textbf{(f)} \mbox{TP}, \mbox{TN}, \mbox{FP}, and \mbox{FN} w.r.t the threshold \textit{k}.}
    \label{fig_readout}
\end{figure*}

\textbf{Cavity-enhanced single-shot readout}\\
After the two swap operations employing the optimal pulses, for designs (a) and (b) we perform cavity enhanced single-shot readout of the SiV$^{-}$ electron spin. Optical readout of solid-state qubits is typically based on resonance fluorescence, and the readout fidelity is limited by the branching ratio corresponding to the spin-flipping transitions. For silicon vacancies, the branching ratio depends on the alignment angle $\alpha$ of the static B-field with the SiV$^-$ symmetry axis. In reference\cite{PhysRevLett.119.223602}, for $\alpha<5^{\circ}$ a cyclicity of more than $\sim10^{5}$ was achieved. Further, despite a low photon collection efficiency, using a laser readout time of 10 ms a single shot readout fidelity of $\sim89\%$ was demonstrated.

Since the duration of the laser readout determines the detection time of our protocol, a low laser readout time is preferable. Since SiV$^-$ is a group-IV CC and is first-order insensitive to the electric field (unlike NV-centers), the readout time can be reduced by several orders of magnitude and the photon collection efficiency can be improved to above $10\%$ by integrating them in nanostructures\cite{Bhaskar2020-ef}.

We consider the architecture as in Fig.~\ref{fig_readout}a, where the SiV$^-$ is in an optical cavity. Fig.\ref{fig_readout}b, shows the $\lambda$-type energy structure corresponding to two of the C-transitions of the SiV$^-$. The parameters $\mu_0$ and $\mu_1$ represent the transition dipole moment of the dipole-allowed spin-conserving and dipole-forbidden spin-flipping optical transitions, respectively. We assume that we are in the SiV strain regime for which $\mu_{1}\ll\mu_{0}$ so that we can neglect the laser induced $\mu_{1}$ transitions. For performing optical readout, the optical cavity is in resonance with the $f_{\uparrow\uparrow^{'}}$ transition and is detuned from the $\downarrow\downarrow^{'}$ transition by $\Delta$. In this regime of operation, the atom-cavity coupling depends on the spin state, which also affects the reflection and transmission coefficients of the cavity. Hence, by probing the transmission/reflection coefficient we can perform a non-destructive single-shot readout of the spin state.

During the readout step, the cavity transmittivity is resonantly probed using a laser pulse of duration $T$, with an incident photon flux of $n_{\ms{pump}}$ (units of number of photons per unit time). In the weak excitation regime (i.e. $n_{\ms{pump}}\ll1/\tau$, where $\tau$ is the modified lifetime of the $\ket{\uparrow'}$ state), the average numbers of transmitted photons when the spin is in each of its two states are given by\cite{Sun_2016, Sun_2018}
\begin{align}
    N_{0}(T) & = \frac{\eta T n_{\ms{pump}}}{(1+C)^{2}}, \;\;\; \mbox{ spin in state $\ket{\uparrow}$}, \label{eq24} \\
    N_{1}(T) & = \eta T n_{\ms{pump}}, \;\;\; \mbox{ spin in state $\ket{\downarrow}$} , \label{eq25} 
\end{align}  
where $\eta$ is the overall photon collection efficiency taking into account the coupling efficiency of the optics, imperfect spatial mode matching between the incident photon and the cavity, and the quantum efficiency of the detector. The atomic cooperativity of the cavity is defined by $C$ = $2g^{2}/(\kappa\gamma)$, where $g$ is the coupling strength between the cavity and the $\mu_0$ transition, $\kappa$ is the cavity decay rate, and $\gamma$ is the decay rate of the excited state. 

The expressions for $N_{0(1)}(T)$ in Eqs.(\ref{eq24}) and (\ref{eq25}) do not take into account that the spin may flip during the measurement process. To take this into account we evolve the master equation for the system during the measurement process and calculate the average value of the photon emission rate when the initial state of the spin is either up or down. We then set $N_{0}(T)$ and $N_{1}(T)$ equal to these average rates, respectively. This gives us the expressions 
\begin{equation}
\label{eq33}
N_{0(1)}(T)=\eta \int_0^T\operatorname{Tr}\left[\kappa \hat{c}^{\dagger}\hat{c} \hat{\rho}_{0(1)}(t)\right] dt,
\end{equation}
where $\hat{c}$ is the annihilation operator for the cavity mode, and $\hat{\rho}_{0(1)}(t)$ is the density matrix of the spin-cavity system at time $t$, for the initial states $\ket{\uparrow}$ and $\ket{\downarrow}$, respectively. We can numerically solve for $\hat{\rho}(t)$ by using a Lindblad master equation, Eq.(\ref{eq2}), using the following readout Hamiltonian in the reference frame w.r.t to the laser frequency $\omega$:
\begin{align}
\label{eq34}
    H_{ro} &= \hbar(\omega_{\ms{c}} - \omega)c^{\dag}c + \hbar(\omega_{\ms{a}} - \omega)\ket{\uparrow'}\bra{\uparrow'} \nn \\
    &+ \hbar(\omega_{\ms{a}} - \Delta-\omega)\ket{\downarrow'}\bra{\downarrow'} + i\hbar g\bigg(c\ket{\uparrow'}\bra{\uparrow} - c^{\dag}\ket{\uparrow}\bra{\uparrow'}\bigg) \nn \\
    &+ i\hbar g\bigg(c\ket{\downarrow'}\bra{\downarrow} - c^{\dag}\ket{\downarrow}\bra{\downarrow'}\bigg) + \hbar\sqrt{\kappa}\epsilon(c + c^{\dagger}).   
\end{align}
To incorporate losses, we use Eq.(\ref{eq3}) for the Lindblad superoperators with a new set of decay parameters 
\begin{align}
  c_{i}^{RO} & \in \{c, \ket{\uparrow}\bra{\uparrow'}, \ket{\downarrow}\bra{\uparrow'}, \ket{\uparrow}\bra{\downarrow'}, \ket{\downarrow}\bra{\downarrow'}, \ket{\uparrow'}\bra{\uparrow'},
\ket{\downarrow'}\bra{\downarrow'}\} , \label{eq111x} \\
  \gamma_{\ms{c}_{i}}^{RO} & \in \{\kappa, \gamma_0, \gamma_1, \gamma_2, \gamma_3, 2\gamma_{\ms{d}}, 2\gamma_{\ms{d}} \}  \label{eq111y}
\end{align}
Here $\gamma_0$ and $\gamma_3$ correspond to the spin-conserving decays, $\gamma_1$ and $\gamma_2$ correspond to the spin-flip decays, and $\gamma_{\ms{d}}$ corresponds to the pure dephasing rate for the states $\ket{\uparrow'}$ and $\ket{\downarrow'}$. The incident field amplitude of the laser, $\epsilon$, is chosen to be much smaller than $g/\sqrt{\kappa}$, so that we stay in the weak field linear regime. 

In order to distinguish between measurement results for $\ket{\uparrow}$ and $\ket{\downarrow}$ we choose a threshold number $k$, comparing the number of collected photons $N_{\ms{c}}$ with $k$. When $N_{\ms{c}}<k$, we report the spin state to be $\ket{\uparrow}$ (presence of a MW photon), whereas when $N_{\ms{c}}>k$ we report the spin state to be $\ket{\downarrow}$ (absence of a MW photon). There are multiple metrics that can be obtained based on the above classification scheme as seen in Fig.~\ref{fig_readout}c. 

\begin{figure}[tbh!]
    \centering
    \includegraphics[width=0.4\textwidth]{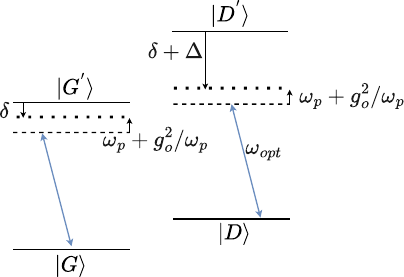}
    \caption{Effective four-level diagram of the ensemble in the dispersive regime.}
    \label{fig_qnd}
\end{figure}

Given two possibilities for the photon (presence or absence) and two measurement results (reporting the presence of absence of the photon) there are four potential outcomes: \mbox{TP} (True Positive, in which we correctly report the presence of the photon); \mbox{TN} (True Negative); \mbox{FP} (False Positive or \enquote{dark count}); \mbox{FN} (False Negative). Assuming that the distribution of the collected photons is a Poissonian, we have the following expressions for the probabilities of the outcomes: 
\begin{align}
\label{eq26}
    \mbox{TN} & = 1 - \sum_{j=0}^{k} \bigg(\frac{\left[N_1(T)\right]^j e^{-N_1(T)}}{j!}\bigg) , \\
\label{eq27}
   \mbox{TP} & = \sum_{j=0}^{k} \bigg(\frac{\left[N_0(T)\right]^j e^{-N_0(T)}}{j!}\bigg) , \\
\label{eq28}
    \mbox{FP} & = \sum_{j=0}^{k} \bigg(\frac{\left[N_1(T)\right]^j e^{-N_1(T)}}{j!}\bigg) , \\
\label{eq29}
    \mbox{FN} & = 1 - \sum_{j=0}^{k} \bigg(\frac{\left[N_0(T)\right]^j e^{-N_0(T)}}{j!}\bigg) . 
\end{align}  
Based on the above expressions, the following figure of merits can be deduced: the success probability $P_{\ms{s}}$, Shannon's mutual information $\textit{I}(X;Y)$, and the intrinsic dark count rate, $\mathcal{D}$. These are given by 
\begin{align}
\label{eq30}
    P_{\ms{s}}(k) & = q\cdot \mbox{TP} + p\cdot \mbox{TN}, \\
\label{eq31}
    \textit{I}(X;Y,k) & = p\cdot \mbox{TN}\cdot\log(\frac{\mbox{TN}}{p\cdot \mbox{TN} + q\cdot \mbox{FN}}) \nn \\
    & \;\;\;\; + p\cdot \mbox{FP}\cdot\log(\frac{\mbox{FP}}{p\cdot \mbox{FP} + q\cdot \mbox{TP}}) \nn \\
    &\;\;\;\; + q\cdot \mbox{FN}\cdot\log(\frac{\mbox{FN}}{p\cdot \mbox{TN} + q\cdot \mbox{FN}}) \nn \\
    &\;\;\;\; + q\cdot \mbox{TP}\cdot\log(\frac{\mbox{TP}}{p\cdot \mbox{FP} + q\cdot \mbox{TP}}), \\
\label{eq32}
    \mathcal{D} & = \mbox{FP}/T_{0} , 
\end{align}
where $q$ and $p$ are the probabilities that the spin is in the states $\ket{\uparrow}$ and $\ket{\downarrow}$,   respectively, \textit{X} and \textit{Y} are the distributions for the spin-state and their predictions respectively, and $T_{0}$ is the time for the complete protocol. As can be seen from Fig.~\ref{fig_readout}e, both $P_{\ms{s}}$ and $\textit{I}(X;Y)$ depend on the threshold parameter $k$. Hence, for each set of parameters, an optimum value of $k$ can be obtained which maximizes $P_{\ms{s}}$ or $\textit{I}(X;Y)$. From Fig. \ref{fig_readout}f, we can see that at optimal value of \textit{k}, $\mbox{TP}$ and $\mbox{TN}$ are high whereas $\mbox{FP}$ and $\mbox{FN}$ are low, as demanded by the detection protocol. Depending on the rarity of the detection event and sensitivity required, different figure of merits can be used. For example for anomaly detection, it has been proposed\cite{6388332} that R\'enyi information is more sensitive than Shannon information due to its asymmetric form. For this paper, we use $P_{\ms{s}}$ and $\textit{I}(X;Y)$ as the figures of merit of our protocol.
\newline\newline
\textbf{Dispersive Readout}\\
For dispersive readout of the excitation stored in the spin ensemble we adapt the scheme introduced in\cite{Wang:17}. As seen from the schematic in Fig.~\ref{fig_qnd}, we consider the energy levels $\ket{G}$ and $\ket{G'}$ with the optical drive of strength $\Omega_{\ms{opt}}$ and frequency $\omega_{\ms{opt}}$. The drive $\Omega_{\ms{opt}}$ is started after the step of ensemble mapping. The parameter $g_{\ms{o}}$ describes the coupling between the state $\ket{G'}$ and phononic cavity. The Hamiltonian of the detection system is given by:
\begin{equation}
    \label{eq35}
    \begin{split}
        \hat{H}_{QND} &= \omega_{\ms{p}}\hat{b}^{\dagger}\hat{b} + \omega_{GG'}\ket{G'}\bra{G'}\\
        & + \frac{\Omega_{\ms{opt}}}{2}(e^{it\omega_{\mbox{\tiny opt}}}\ket{G}\bra{G'} + e^{-it\omega_{\mbox{\tiny opt}}}\ket{G'}\bra{G})\\
        &+ g_{\ms{o}}(\hat{b} + \hat{b}^{\dagger})\ket{G'}\bra{G'}.
    \end{split}
\end{equation}
where, $\omega_{GG'} = \omega_{\ms{opt}} + \omega_{\ms{p}} + g_{\ms{o}}^2/\omega_{\ms{p}} + \delta$. After applying the Schrieffer-Wolff transformation w.r.t the unitary 
\begin{align}
    U = \exp\left[ \frac{g_{\ms{o}}}{\omega_{\ms{p}}}(\hat{b}^\dagger - \hat{b})\ket{G'}\bra{G'}  \right]
\end{align} 
and a sequence of two rotating-wave approximations, first w.r.t. 
\begin{align}
    \hat{H}_{r1} = \omega_{\ms{p}}\hat{b}^{\dagger}\hat{b} + \left(\omega_{GG'} - \frac{g_{\ms{o}}^{2}}{\omega_{\ms{p}}} - \delta \right) \ket{G'}\bra{G}
\end{align}
and then w.r.t. $\hat{H}_{r2} = \delta\ket{G'}\bra{G'}$, one obtains the following Hamiltonian:
\begin{equation}
    \label{eq36}
    \begin{split}
      \hat{H'} = \frac{g_{\ms{o}}\Omega_{\ms{opt}}}{2\omega_{\ms{p}}}(e^{i\delta t}\hat{b}^{\dagger}\ket{G'}\bra{G} + e^{-i\delta t}\hat{b}\ket{G}\bra{G'}).        
    \end{split}
\end{equation}
Finally, in the dispersive regime the application of a perturbation expansion in which the interaction with the phonon mode perturbs the spin ensemble, the effective Hamiltonian becomes 
\begin{equation}
    \label{eq37}
    \begin{split}
    \hat{H}_{DR} &= \frac{g_{\ms{o}}^{2}\Omega_{\ms{opt}}^{2}}{4\omega_{\ms{p}}^{2}\delta} \bigl[ \hat{b}^{\dagger}\hat{b}(\ket{G'}\bra{G'} - \ket{G}\bra{G})
    + \ket{G}\bra{G} \bigr]. 
    \end{split}
\end{equation}
There are three stages to the dispersive readout. In the first stage a coherent drive is applied to the phononic cavity to prepare the cavity mode in a coherent state. The cavity photon number should be as large as possible but no larger than 
\begin{align}
  n_{crit} = \frac{4\omega_{\ms{p}}^{2}\delta^{2}}{g_{\ms{o}}^{2}\Omega_{\ms{opt}}^{2}}
\end{align}
to maintain the validity of the perturbation expansion. The phononic cavity decay rate $\gamma_{\ms{ph}}$ also determines its coupling to the drive. Fast readout requires a large value of $\gamma_{\ms{ph}}$ which also increases the rate of phonon loss. In the second stage phonons interact with the spin-ensemble and the excitation in the ensemble produces a phase shift of the coherent state. The rate at which this phase shift is accumulated is given by the effective coupling rate 
\begin{align}
  2\chi = \frac{g_{\ms{o}}^{2}\Omega_{\ms{opt}}^{2}}{2\omega_{\ms{p}}^{2}\delta}.  
\end{align}
In the third stage heterodyne detection is used to measure the phase and amplitude of the output from the cavity. 

\subsection{Numerical simulation}
We now present the results of simulating the operation of the three types of photon-detector. As seen above, the protocols for each photon-detector have two parts: Mapping (employing one of the protocols QST, AdM, or EnM) and Readout (either SSR or DR). The protocols have two metrics, the success probability $P_{\ms{s}}$ and the mutual information \textit{I(X;Y)}. 
 
\subsubsection{Quantum State Transduction}
\begin{figure*}[hbt!]
    \centering
    \includegraphics[width=0.7\textwidth]{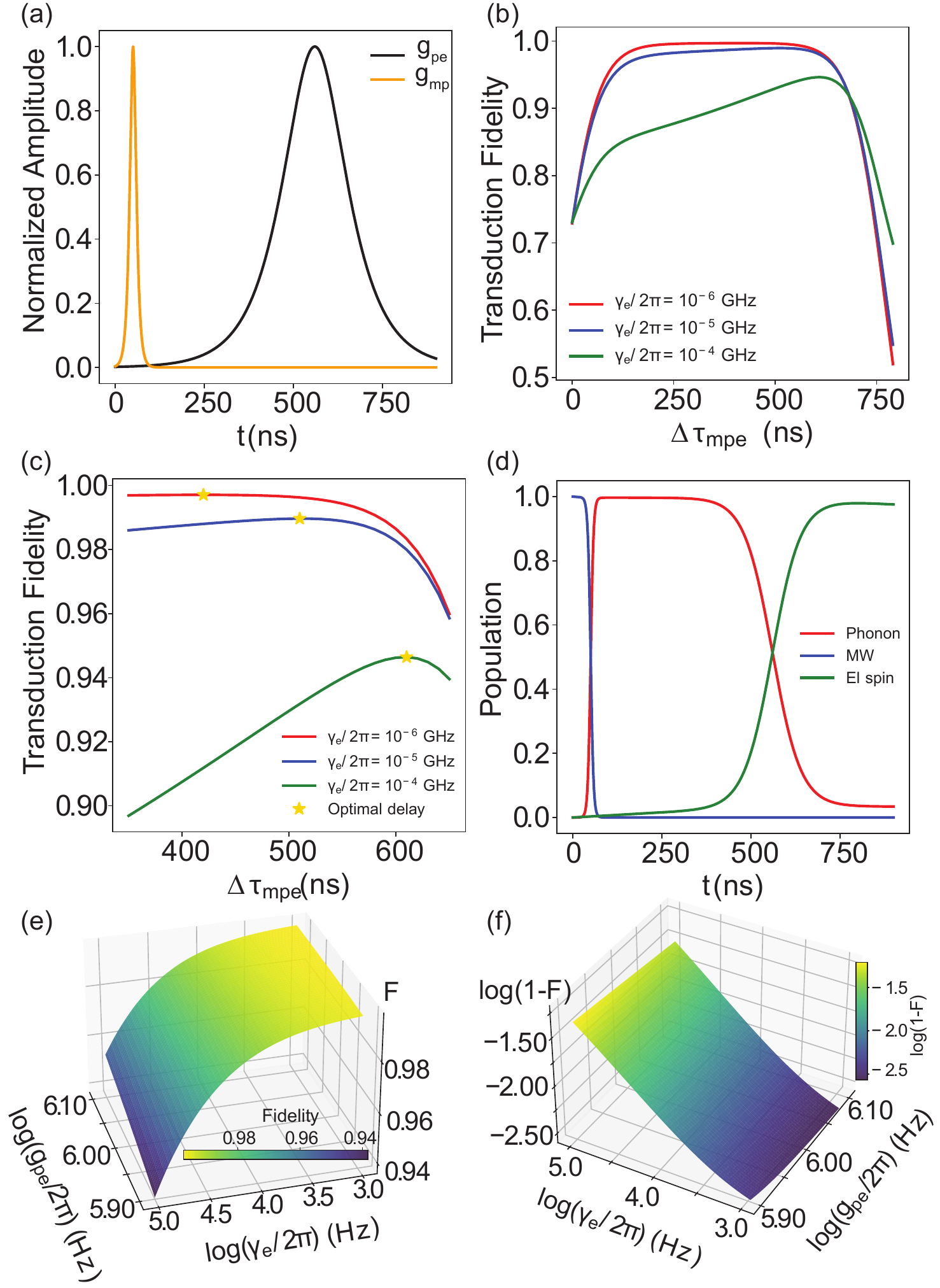}
    \caption{\textbf{State transfer from MW cavity to electron spin.} \textbf{(a)} Time-dependent pulse sequences. \textbf{(b)} Transduction fidelity as a function of inter-pulse time-delay, with further zoomed in version in \textbf{(c)}, \textbf{(d)} time dependent population for the MW, phonon, electron spin. \textbf{(e), (f)} variation of infidelity with $g_{\ms{pe}}$ and $\gamma_{\ms{e}}$.}
    \label{fig_qst}
\end{figure*}

We simulate the master equation with the Hamiltonian described in Eq.(\ref{eq1}) and collapse operators given in Eq.(\ref{eq3}). Fig.~\ref{fig_qst} shows the results of this simulation using the time-dependent pulse sequences for the MW-phonon swap and the phonon-spin swap (Fig.~\ref{fig_qst}a). As the time delay between the two pulses is varied, the transduction fidelity changes, as shown in Fig.~\ref{fig_qst}b (Fig.~\ref{fig_qst}c provides a zoom-in). The point at which the maximum fidelity is obtained is marked by the yellow star. Also as $\gamma_{\ms{e}}$ is reduced, the fidelity increases and the optimal time-delay decreases. Using this optimal value of time-delay and $\gamma_{\ms{e}}/2\pi$ = $10$ kHz, we plot the time dependent population for the MW, phonon, and  electron spin modes in Fig.~\ref{fig_qst}d. Next, in Fig.~\ref{fig_qst}e(f), we vary $g_{\ms{pe}}$ and $\gamma_{\ms{e}}$ over a realistic range to observe the variation in the fidelity (infidelity). For each pair of parameters [$g_{\ms{pe}}$, $\gamma_{\ms{e}}$], we adjust the time delay to maximize $\mathcal{F}$. We observe that $\mathcal{F}$ varies in the range (0.937, 0.997) with a standard deviation of 0.015.

\subsubsection{Adiabatic Mapping}
\begin{figure*}[hbt!]
    \centering
    \includegraphics[width=0.8\textwidth]{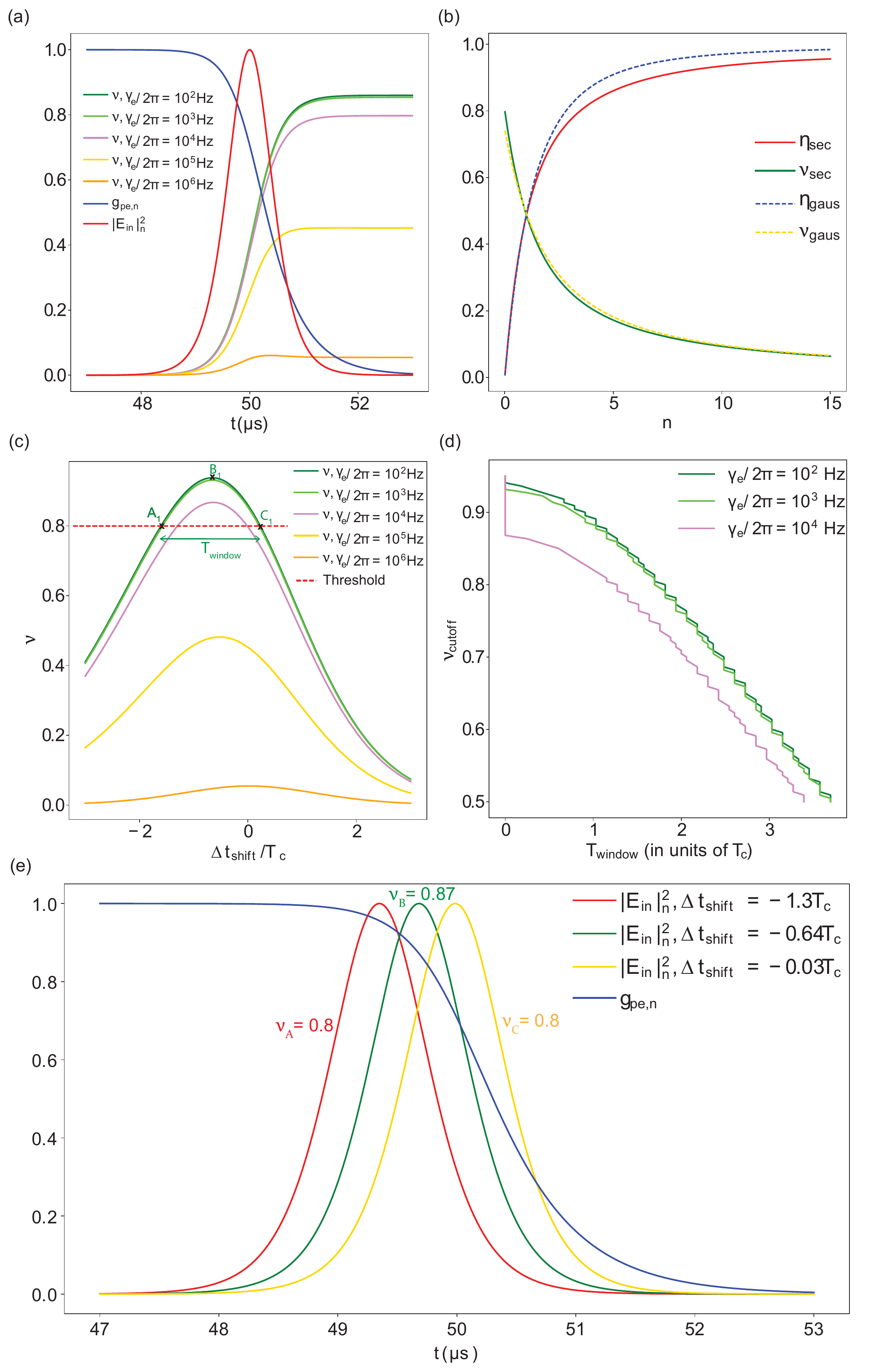}
    \caption{\textbf{Adiabatic Mapping from incoming photon wavepacket to electron spin.} \textbf{(a)} Normalized time profile of the incoming weak coherent pulse (n = 0.01), optimum drive $g_{pe,n}$ and the transfer fidelity $\nu$. \textbf{(b)} Variation of $\nu$ and efficiency $\eta$ with the mean photon number $n$. \textbf{(c)} Variation in $\nu$ with $t_{shift}$. Red dashed line denotes the threshold. \textbf{(d)} $\nu_{cutoff}$ vs $T_{window}$,  \textbf{(e)} Temporal profile w.r.t. the drive $g_{\ms{pe}}$ for the three points mentioned in (\textbf{(c)}).} 
    \label{fig_adm}
\end{figure*}

We simulate the master equation with Hamiltonian and optimal driving described in Eqs.(\ref{eq7})-(\ref{eq10}) and (\ref{eq16}). Fig.~\ref{fig_adm}a shows the normalized time profile of the incoming weak coherent pulse (n = 0.01), with the normalized optimum drive $g_{pe,n}$ and the transfer fidelity $\nu$ which saturates at 0.86. Fig.~\ref{fig_adm}b shows the variation of the fidelity $\nu$ and efficiency $\eta$ with the mean photon number $n$ in the incoming wavepacket. For low photon number, these figures of merit coincide for two different temporal shapes of the incoming wavepacket (hyperbolic secant and Gaussian). Shifting the incoming wavepacket by $t_{shift}$ from the center of the drive leads to a variation in the fidelity, with a maximum of 0.939 achieved for point B, as seen in Fig.~\ref{fig_adm}c. Fixing the cutoff fidelity $\nu_{cutoff}$ at 0.8 gives a range (A-C) with a detection window $T_{window}$. Changing $\nu_{cutoff}$ changes the detection window, implying that lowering the cutoff fidelity increases the this window (Fig.~\ref{fig_adm}d). From Fig.~\ref{fig_adm}e we see the temporal profile w.r.t. the drive $g_{\ms{pe}}$ for the three points mentioned in (\textbf{(c)}), giving a window of $\sim 1.82 T_{\ms{c}}$.

\subsubsection{Ensemble Mapping}
\begin{figure*}[hbt!]
    \centering
    \includegraphics[width=0.8\textwidth]{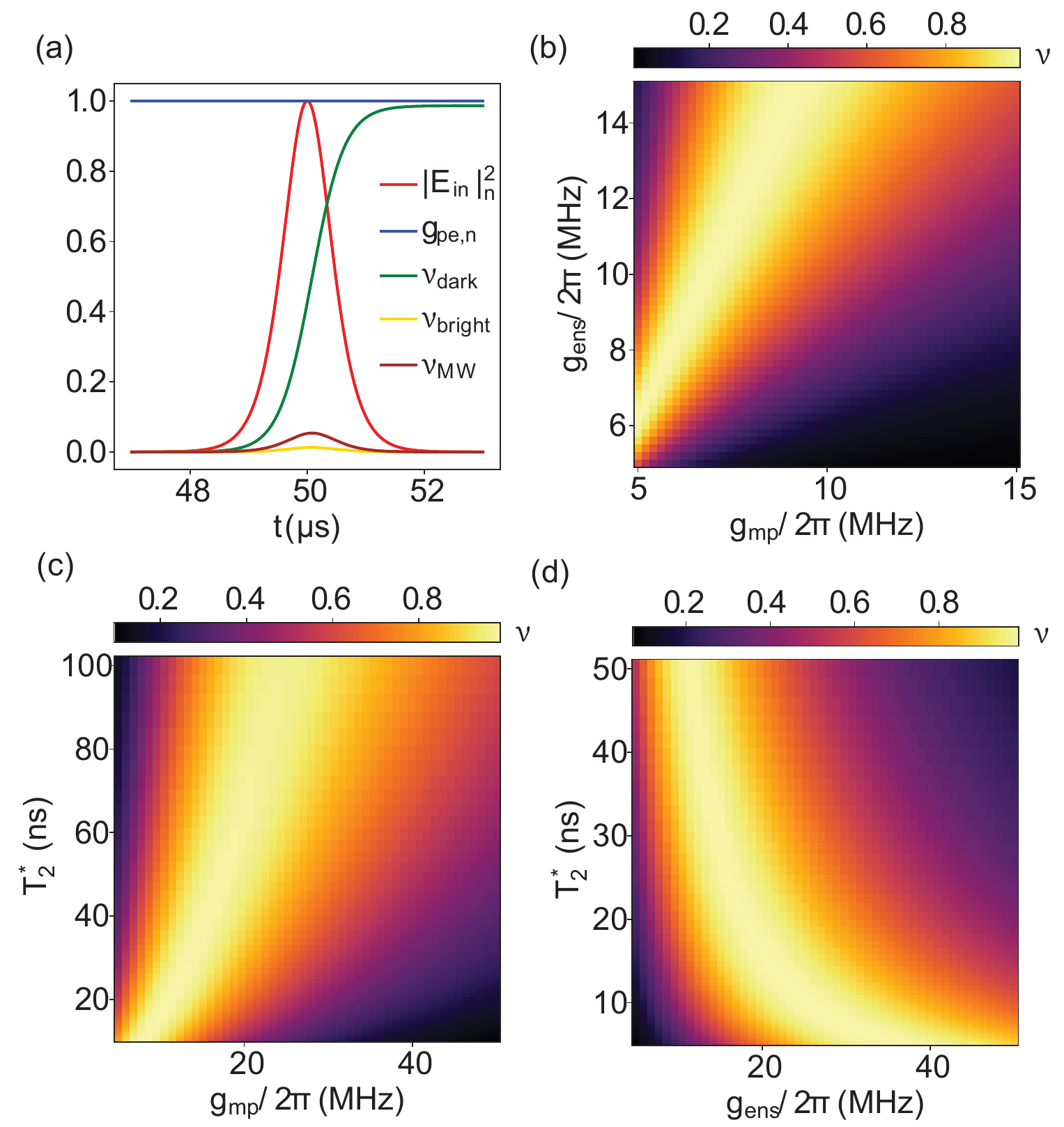}
    \caption{\textbf{Ensemble Mapping from incoming photon wavepacket to electron spin ensemble}. \textbf{(a)} Normalized time profile of the incoming weak coherent pulse (n = 0.01), drive $g_{pe,n}$ and the transfer fidelity $\nu_{dark}$. \textbf{(b)}, \textbf{(c)}, and \textbf{(d)} color maps with varying $g_{\ms{mp}}$, $g_{\ms{ens}}$ and $T^{*}_{2}$.} 
    \label{fig_ens}
\end{figure*}

We simulate the master equation with the Hamiltonian given in Eq.(\ref{eq19}). In Fig.~\ref{fig_ens}a we show the normalized time profile of the incoming weak coherent pulse (n = 0.01), with the constant normalized drive $g_{pe,n}$, and the transfer fidelity $\nu_{dark}$ which saturates at 0.986. Fig.~\ref{fig_ens}b, c, and d show the result of  varying $g_{\ms{mp}}$, $g_{\ms{ens}}$, and $T^{*}_{2}$, respectively. These figures can be used to find a suitable range of coupling strengths which maximize the transfer fidelity $\nu_{dark}$.

\subsubsection{Single-shot readout}
In Fig.~\ref{fig_readout}d we show $P_{\ms{s}}$ as a function of the laser readout time $T$ and cooperativity $C$, based purely on the expressions for $N_0(T)$ and $N_1(T)$ given in Eqs.(\ref{eq24}) and (\ref{eq25}). %\EA{QUESTION: What's with the plateaus in the $C=0.5$ case (also seen at the end of 5 and 10?}%. 
In this case $P_{\ms{s}}$ increases monotonically with $T$, but as discussed above this simple picture is incomplete as it does not include the possibility of spin-flips during the readout time. To obtain a more accurate simulation of the measurement we must simulate the master equation with the Hamiltonian in Eq.(\ref{eq34}) and decay operators given in Eqs.(\ref{eq111x}) and (\ref{eq111y}). This provides us with the average values of $N_{0(1)}$ given that the spin is  initially in the states $\ket{\uparrow}(\ket{\downarrow})$, and subsequently the readout metrics in Eqs.(\ref{eq26})-(\ref{eq31}). Fig.~\ref{fig_ro}a shows the variation of $P_{\ms{s}}$ and $I(X;Y)$ with the laser readout time. These initially increase and then start to reduce due to spin-flip transitions and other decay channels, with the orange stars marking the optimum point at $T\sim240$ ns. Taking the optimum value of $T$, we vary the efficiency $\eta$ in Fig.~\ref{fig_ro}b, and the detuning $\Delta$ in Fig.~\ref{fig_ro}c, showing an asymptotic behavior for both. Using the optimal value of $T$ and $\eta=0.85$, we then run a quantum Monte-Carlo simulation with the same Hamiltonian to obtain a histogram shown in Fig.~\ref{fig_ro}d. Here the dotted line gives the thresholding parameter, such that only an event falling to its left is classified as a photon-detection. The resulting overall success probability is $\sim0.972$. 

\subsubsection{Dispersive Readout}
\begin{figure*}[hbt!]
    \centering
    \includegraphics[width=0.7\textwidth]{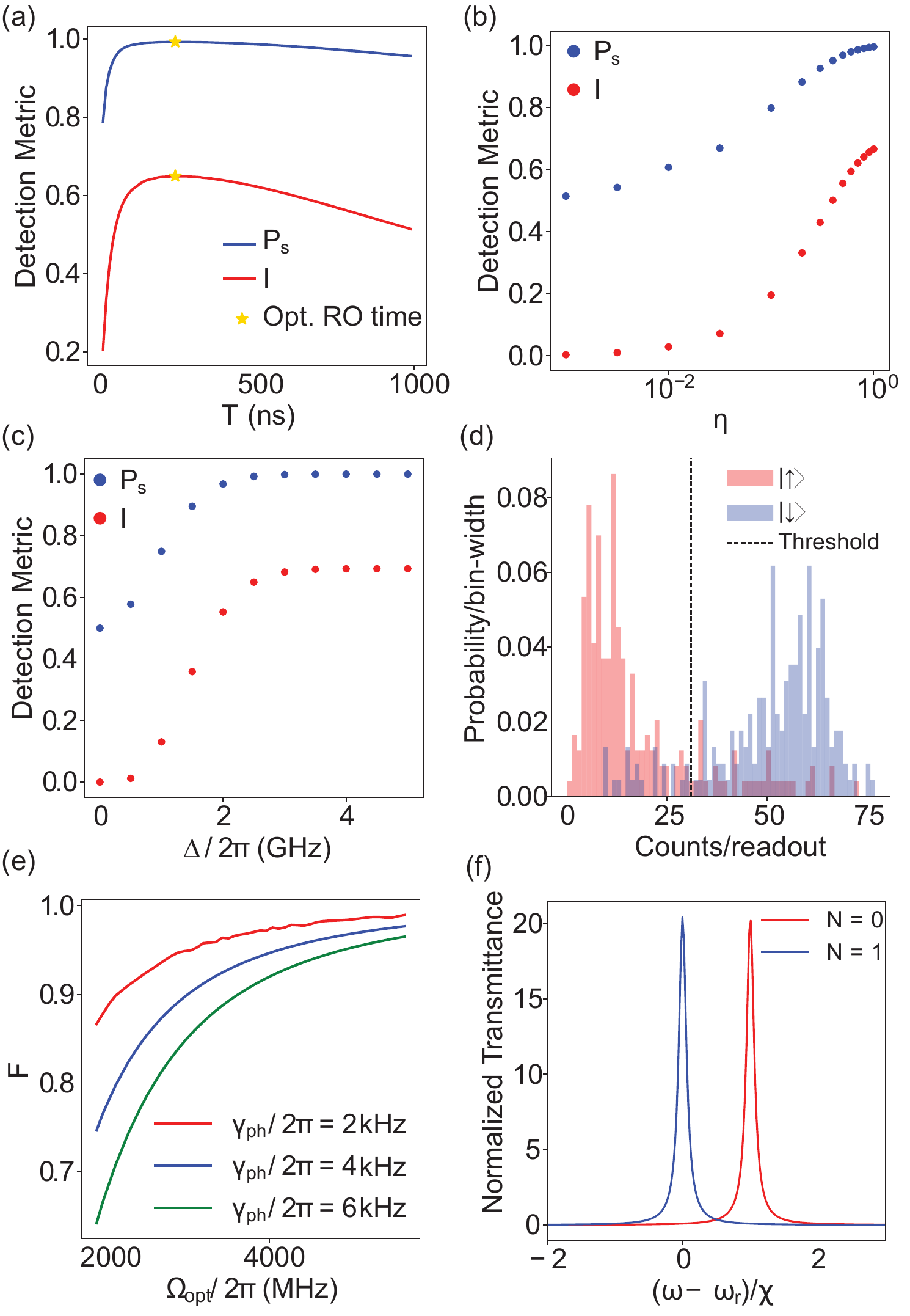}
    \caption{\textbf{Readout of electron spin.} \textbf{(Single-shot)} \textbf{(a)}, \textbf{(b)}, \textbf{(c)} Variation of $P_{\ms{s}}$ and $I(X;Y)$ with laser readout time $T$, collection efficiency $\eta$ and detuning $\Delta$, \textbf{(d)} histogram plot with  detector click statistics for two states $\ket{\uparrow}$ and $\ket{\downarrow}$ and the dotted line shows the thresholding. \textbf{(Dispersive)} \textbf{(e)} Variation of dispersive readout fidelity $\mathcal{F}$ with optical drive strength $\Omega_{\ms{opt}}$, \textbf{(f)} Shift in resonance frequency of the phononic resonator depending on the MW photon state.}
    \label{fig_ro}
\end{figure*}

We simulation the master equation with the Hamiltonian in Eq.(\ref{eq37}) to simulate the dispersive readout. Fig.~\ref{fig_ro}e shows the variation of the readout fidelity $\mathcal{F}$ with optical drive strength $\Omega_{\ms{opt}}$. Fig.~\ref{fig_ro}f shows the shift in the resonance frequency of the phononic resonator due to the presence of a photon, which yields a fidelity of 0.924.

\begin{figure*}[htb!]
    \centering
    \includegraphics[width=0.9\textwidth]{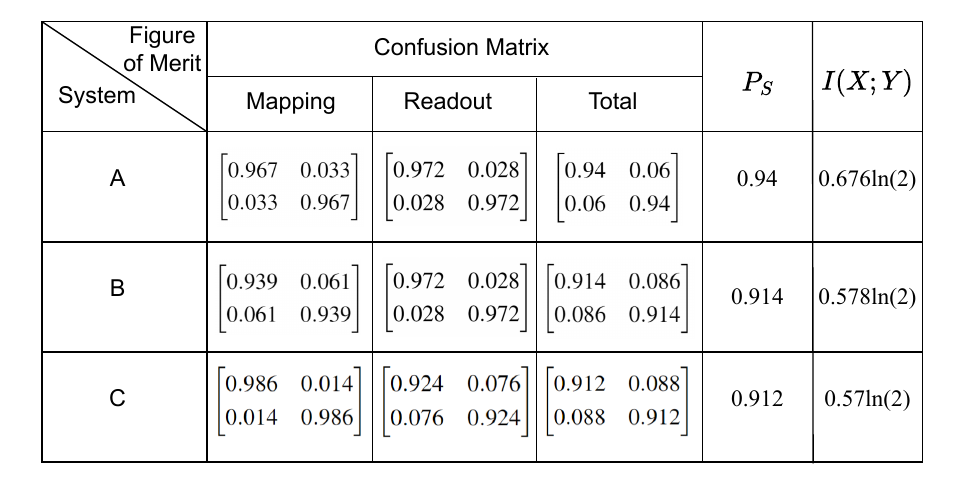}
    \caption{Simulation Summary}
    \label{summary}
\end{figure*}

\section{Discussions}
Now that we have simulated the performance of the individual components, we can determine the confusion matrices for these components as shown in Fig.~\ref{summary} and combine the two parts of each protocol. The overall confusion matrix for each detector is the product of the confusion matrices for the mapping and readout steps. For each of the detectors (a), (b), and (c), we obtain a maximum detection efficiency of 0.94, 0.914, and 0.912, respectively. 

We have shown that using mechanics to mediate an interaction between color centers and itinerant microwave photons has the near-term potential to enable high-efficiency microwave photon detection. We have explicitly evaluated a design that uses this mechanism to realize a high efficiency photon-detector in the traditional sense. We have also presented and evaluated designs for single-photon detectors for more specialized applications: (a) a detector that determines the presence or absence of a photon in a microwave cavity, and (b) a detector that determines the presence of an incident traveling-wave photon when the pulse shape and time of arrival of the photon are known. Using experimental parameters from platforms that have already been performed, the detection efficiencies we obtain are in the range $0.91-0.94$, and the corresponding mutual information for the detectors is $0.57\ln(2)-0.67\ln(2)$. These efficiencies are larger than those of the various alternatives currently available\cite{Gu2017-di, PhysRevLett.102.173602, PhysRevA.84.063834, balembois2023practical, PhysRevLett.107.217401, Sathyamoorthy2016-ge, Inomata2016-he, Woodman2023-xk}. There exist platforms such as Ref. \cite{balembois2023practical} which have a lower dark count rate of $85 s^{-1}$, but these suffer from a low detection efficiency of $0.43$. Hence, our proposal has the potential to significantly impact microwave photon-detection technology.

\section{Code Availability}
The simulations were performed using QuTiP\cite{qutipQuTiPQuantum} Python and the codes can be found in our GitHub repository\cite{githubGitHubPanand2257SinglePhotonDetection}.

\section{Acknowledgements}
P.A. would like to thank Amit and Deepali Sinha Foundation Fellowship from MIT. E.G.A. recognizes support from the Army Educational Outreach Program Postdoctoral Fellowship.

\section{Author contributions}
D.R.E., P.A., and E.G.A conceived the project. P.A. performed the simulations, and optimal control modeling. K.J. and E.G.A. assisted in simulations. P.A., E.G.A., and K.J. prepared the manuscript. All
authors discussed results and revised the manuscript. D.R.E. supervised the project.

\section{Competing interests}
The authors declare no competing interests.

\section{Additional information}
For our simulations we use parameter values corresponding to previously demonstrated platforms. The set of simulation parameters for the five parts is presented in Tables 1-4 in the Supplementary Information here\cite{googleSupplementsGoogle}.

\newpage
\bibliography{MWSPD}

\end{document}